\documentclass[superscriptaddress,nofootinbib,preprint,showpacs]{revtex4}
\usepackage[latin1]{inputenc}
\usepackage{graphicx}
\usepackage{amssymb}
\usepackage{color}
\usepackage{float}
\usepackage{amsmath}
\usepackage{amsfonts}
\usepackage{dcolumn}
\usepackage{hyperref}
\usepackage{amsthm}
\usepackage{color}
\usepackage{bm}

\def\3nab{\tilde{\nabla}}

\def\be {\begin{equation}}
\def\ee {\end{equation}}
\def\bq {\begin{eqnarray}}
\def\eq {\end{eqnarray}}
\def\bs {\begin{subequations}}
\def\es {\end{subequations}}

\newcommand{\barray}{\begin{array}}
\newcommand{\earray}{\end{array}}

\begin{document}

\title{Relativistic stellar collapse of initially static configurations}
\author{Keshlan S. Govinder}
\email{govinder@ukzn.ac.za}
\affiliation{Astrophysics Research Centre, Discipline of Mathematics, University of KwaZulu-Natal, Private Bag X54001, Durban 4000, South Africa}

\author{Megandhren Govender}\email[]{megandhreng@dut.ac.za}
\affiliation{Department of Mathematics, Faculty of Applied Sciences, Durban University of Technology, P O Box 1334, Durban 4000, South Africa}

 \author{Sunil D. Maharaj}\email[]{maharaj@ukzn.ac.za}
\affiliation{Astrophysics Research Centre, Discipline of Mathematics, University of KwaZulu-Natal, Private Bag X54001, Durban 4000, South Africa}

\begin{abstract}
We study the dynamics of a radiating relativistic star in the presence of the electromagnetic field and the cosmological constant. The evolution of the model originates from an initially static configuration. The temporal behaviour of the system is governed by a nonlinear second order equation which is studied using a phase plane analysis. We show that in the general case, with all physical parameters present, fold bifurcations arise which are related to nonzero charge and cosmological constant. Our treatment clarifies earlier results and we extend our study to the general solution space. We obtain a more comprehensive understanding of the temporal evolution for a shear-free configuration during collapse.
 \vspace{2cm}
 
 \end{abstract}  

 \maketitle

\section{Introduction}

The problem of gravitational collapse of self-gravitating stellar objects has occupied the interest of researchers since the proposal of ``dark bodies" by Pierre-Simon Laplace and John Michell in the late 1700's. In 1931 Chandrasekhar showed that there was an upper limit to the mass of white dwarfs which ensured equilibrium \cite{chandra}. Subsequent work on stars with masses above the Chandrasekhar limit showed that such objects will undergo continued gravitational collapse. This inspired the seminal work of Oppenheimer and Snyder \cite{os39} in which they investigated the collapse of a dust sphere and provided the first insights into the possible end state of a dying star. The singularity theorems of Penrose are regarded as the most fundamental contributions to understanding gravitational collapse and the ultimate fate of stars within  the framework of Einstein's general relativity. Penrose's Cosmic Censorship Conjecture  asserts that any reasonable matter configuration collapsing under the influence of gravity will ultimately meet its fate as a black hole \cite{Penrose}. The idea of singularities and trapped surfaces were developed further by Hawking and Penrose. The existence of naked singularities has led to various enunciations of the Cosmic Censorship Conjecture which calls into question the nature of matter making up the collapsing body.  This has prompted researchers to consider physically viable stellar models in which the matter content includes pressure anisotropy, shear viscosity, bulk viscosity, electric charge amongst others. 

With the discovery of the Vaidya solution in 1951 \cite{vaidya}, it became possible to study the gravitational collapse of stars dissipating energy in the form of heat flux. In his study of nonadiabatic gravitational collapse of radiating spheres, Santos \cite{santos} derived the junction conditions required for the smooth matching of the interior spacetime to the outgoing Vaidya metric. The matching conditions require the nonvanishing of the pressure on the boundary of the collapsing sphere which arises from the conservation of momentum across the comoving hypersurface. This boundary condition encodes the temporal behaviour of the radiating body. The study of collapsing stars and dissipating energy in the form of a radial heat flux to the Vaidya exterior has been an active and fruitful area of research since the early 1980's. Seminal work by Herrera and co-workers on dynamical stability \cite{chan1,chan2}, the role of anisotropy \cite{santos1} and shear \cite{euclid,chan3,free} have revealed new insights into possible outcomes of gravitational collapse. The evolution of temperature and luminosity profiles were studied extensively by Maharaj and collaborators within the framework of extended causal thermodynamics \cite{mart2,mg05}. They showed that relaxational effects lead to higher core temperatures and cannot be ignored especially towards the late stages of collapse.   
 
 The collapse from an initially static configuration has been widely explored in the literature and has provided rich insights into the time evolution of the thermodynamical variables as the fluid loses hydrostatic equilibrium. The framework to study such a scenario was first presented by de Oliveira {\it et al} \cite{doks1} and revisited by the same authors who explored further the end states of such a collapse \cite{doks}. This model has been extended to include charge \cite{jaryl11}, anisotropy \cite{jaryl1}, shear \cite{wchan,rmk}, equation of state \cite{bog} and higher dimensions \cite{bhui}. 

A systematic study of shear-free collapse was undertaken by Maharaj and Govinder \cite{mg2025}. They investigated the asymptotic behaviour of the stellar boundary for neutral fluids as well as charged fluids. Each scenario was further analysed in the presence of the cosmological constant. It was also shown that the presence of charge and the cosmological constant fundamentally affects the temporal behaviour of the interior stellar spacetime. In that treatment, only separability of the metric potentials was assumed leading to a nonlinear differential equation that could be analysed using a dynamical systems analysis.

Stellar collapse has also been studied in higher dimensions. In the general relativity scenario, it was shown that dimension affects the  physical features of the models \cite{mbgg2025}. A similar observation was made when the Einstein-Gauss-Bonnet corrections to general relativity are considered \cite{egbsub}. An asymptotic analysis shows that the inclusion of the Gauss-Bonnet terms leads to behaviour which is different to general relativity.

Given the success of asymptotic analysis for the evolution of stellar collapse, here we apply the method to the collapse of an initially static configuration. We should how this approach provides a more complete understanding of stellar temporal evolution that previously presented. We also unpack the role of charge and the cosmological constant in stellar collapse. The metric functions are separable similar to \cite{mg2025} but there is the additional requirement that the model is {\it initially static}. The condition of initial staticity leads to governing dynamical equations which are different from the analysis of Maharaj and Govinder \cite{mg2025} or other dynamical systems analyses. The subsequent dynamical systems study is qualitatively different and leads to new physical features revealing new dynamical behaviour.

\section{Model}
For the interior of the star, we start with the general metric for a shear-free spherically symmetric spacetime in isotropic coordinates $(t,r,\theta,\phi)$ so that
\be ds^2 = - A^2(r,t) dt^2 + B^2(r,t) \left[dr^2+r^2 (d\theta^2+\sin^2\theta d\phi^2)\right], \label{int1} \ee
with the energy momentum tensor 
\be T_{ab} = (\rho+p_r) u_au_b + p_t g_{ab} + (p_r-p_t)X_a X_b + q_a u_b + q_b u_a,\ee 
where $\rho$ is the energy density, $p_r$ is the radial pressure,  $p_t$ is the tangential pressure and $q^a=(0,q,0,0)$ is the heat flow vector. (Here $q$ is the magnitude of the heat flow.) These quantities are measured relative to the fluid $4$--velocity $u^a$ with $u^au_a=-1, X^aX_a=1, u^aX_a=0$ and $q^au_a=0.$ The expansion, $\Theta$, is given by
\be \Theta = \frac{3\dot{B}}{AB}\, . \label{expansion} \ee 

The resulting Einstein field equations are then
\bs \label{efe1}
\bq
8 \pi \mu &=& \frac{3\dot{B}^2}{A^2B^2} - \frac{1}{B^2} \left(2\frac{B''}{B}-\frac{B'{}^2}{B^2} + \frac{4B'}{rB}\right),\label{efe1a} \\
8\pi p_r &=& \frac{1}{A^2}\left(-2\frac{\ddot{B}}{B}-\frac{\dot{B}^2}{B^2} + 2 \frac{\dot{A}\dot{B}}{AB}\right) \nonumber \\
&&\mbox{} + \frac{1}{B^2}\left(\frac{B'{}^2}{B^2}+2\frac{A'B'}{AB} + \frac{2A'}{rA}+\frac{2B'}{rB}\right), \label{efe1b} \\
8\pi p_t &=& \frac{1}{A^2} \left(-2\frac{\ddot{B}}{B} - \frac{\dot{B}^2}{B^2} + 2\frac{\dot{A}}{A}\frac{\dot{B}}{B}  \right) \nonumber \\ &&\mbox{}  + \frac{1}{B^2} \left(\frac{A''}{A}+\frac{B''}{B}+\frac{B'}{rB} +\frac{A'}{rA}-\frac{B'{}^2}{B^2}\right), \label{efe1c} \\
8 \pi q &=& -\frac{2}{AB^2} \left(-\frac{\dot{B}'}{B}+\frac{B'\dot{B}}{B^2} + \frac{A'\dot{B}}{AB}\right), \label{efe1d} 
\eq \es
where the dots and primes represent differentiation with respect to $t$ and $r$ respectively and we use units in which $G=c=1$. 

We now match the interior metric (\ref{int1}) to the exterior Vaidya metric \cite{vaidya}
\bq ds^2 &=& -\left(1-\frac{2m(v)}{\sf r} \right)d v^2 - 2 dv d{\sf r} + {\sf r^2} (d \theta^2+\sin^2\theta^2 d \phi^2), \label{s4} \eq
where $m(v)$  is the mass of the star observed at infinity, and the exterior spacetime coordinates are $(v,{\sf r},\theta, \phi)$, at the boundary $r=R_\Sigma$.  This requirement,  together with the matching of the extrinsic curvatures, requires \cite{santos}
\be\left. (p_r=qB)\right|_{r=r_\Sigma}.\label{match} \ee

We now choose the potentials for the interior metric to be
\be A = A_0(r), \qquad B = B_0(r) f(t), \label{pot} \ee
where $f(t)$ is positive and the radial functions describe an initial static perfect fluid. In our model, time starts at some initial $t=t_0$ and progresses until the formation of the event horizon.

 The system (\ref{efe1}) then becomes
 \bs \label{efe2}
\bq
8 \pi \mu &=& \frac{1}{f^2}\left(8\pi \mu_0+\frac{3}{A_0^2}\dot{f}^2\right),\label{efe2a} \\
8\pi p_r &=& \frac{1}{f^2}\left(8\pi (p_r)_0 - \frac{1}{A_0^2}\left(2 f \ddot{f} +\dot{f}^2\right)\right), \label{efe2b} \\
8 \pi p_t &=& \frac{1}{f^2}\left(8\pi (p_t)_0 - \frac{1}{A_0^2}\left(2 f \ddot{f} +\dot{f}^2\right)\right), \label{efe2c} \\
8 \pi q &=& -\frac{2A_0'}{A_0^2B_0^2}\frac{\dot{f}}{f^3},\label{efe2d} 
\eq \es
where $\mu_0$,  $(p_r)_0$ and $(p_t)_0$ are the static density, radial pressure and tangential pressure respectively. We note that, if the initial static state is isotropic, ie. $(p_r)_0=(p_t)_0$, then the pressure is always isotropic, ie.  $p_r=p_t$ from (\ref{efe2b}) and (\ref{efe2c}) $\forall\,t$. However, if the initial static state is anisotropic, then the pressure will remain anisotropic.

Taking (\ref{pot}) into account, (\ref{match}) now becomes (via (\ref{efe2}))
\be 2 f \ddot{f} + \dot{f}^2 - 2 a \dot{f} = 0, \label{bound1} \ee
where  $(p_r)_0=0$ on the stellar boundary and $a=A_0'/B_0>0$. 

If we replace the time-dependent terms in (\ref{efe2b}) via (\ref{bound1}) and note that $p_r$ is necessarily positive, then  the system undergoes contraction since
\be \dot{f} \leq 0. \label{fdotcons} \ee
(See also (\ref{expansion}) for the metric (\ref{int1}) when (\ref{pot}) is imposed.) 
Thus, the evolution of initial static configurations must result in collapse; expansion is not allowed.

An invariant of (\ref{bound1}) is given by
\be \dot{f} = -2a\left(\frac{b}{\sqrt{f}}-1\right), \label{firstint} \ee
where $b$ is an arbitrary constant of integration (and technically is the label for the invariant/first integral).
An implicit solution of (\ref{bound1}) is given by
\be t + c = \frac{1}{a}\left[\frac12 f + b \sqrt{f}+b^2 \log\left(1-\frac{\sqrt{f}}{b}\right)\right], \label{impsol} \ee
where $c$ is an arbitrary constant of integration (which can removed due to the translational invariance of (\ref{bound1})). While it can be useful to write down such (quantitative) solutions, interpreting the behaviour of (\ref{bound1}) via (\ref{impsol}) is difficult and, indeed, can be misleading. In the next section, we analyse (\ref{bound1}) from a qualitative perspective.

 The luminosity perceived by an observer placed at infinity from the collapsing body is given by
\be \label{lum}
L_{\infty} = 2a^2\left(\frac{rB_0}{A_0}\right)^2\left(\frac{b-\sqrt{f}}{\sqrt{f}}\right)\left(\frac{1}{1 + z_\Sigma}\right)^2,\ee where the surface redshift, $z_\Sigma$ is given by
\be \label{Z}
z_{\Sigma} = \left[\frac{r(rB_0)' + 2ar^2B_0^2A_0^{-1}\sqrt{f}(b - \sqrt{f})}{rB_0f - 2m}\right]_\Sigma - 1,
\ee
from which we note that $L_\infty \rightarrow 0$ for $(rB_0f) \rightarrow 2m$ {\it{or}} $f(t) \rightarrow b$, signifying the time of formation of the horizon. For a full discussion of the luminosity profile, see \cite{doks}.

 \section{Analysis}
 
We  investigate the asymptotic behaviour of (\ref{bound1}) via a phase plane approach \cite{strog}. We rewrite (\ref{bound1}) as the system
  \bs \label{ds1}
\bq
\dot{f}&=&y, \label{ds1a} \\
2 f \dot{y} &=& - y^2+2 a y, \label{ds1b}
 \eq \es
 noting that $f=0$ is a singularity in (\ref{ds1b}) which means that trajectories are  confined to either the left-half-plane or the right-half-plane. This indicates that if $f(t_0)>0$ then $f(t)>0\ \forall\ t \in \mathbb{R}$ and vice versa.
 
\begin{figure}[t]
\includegraphics[width=5in]{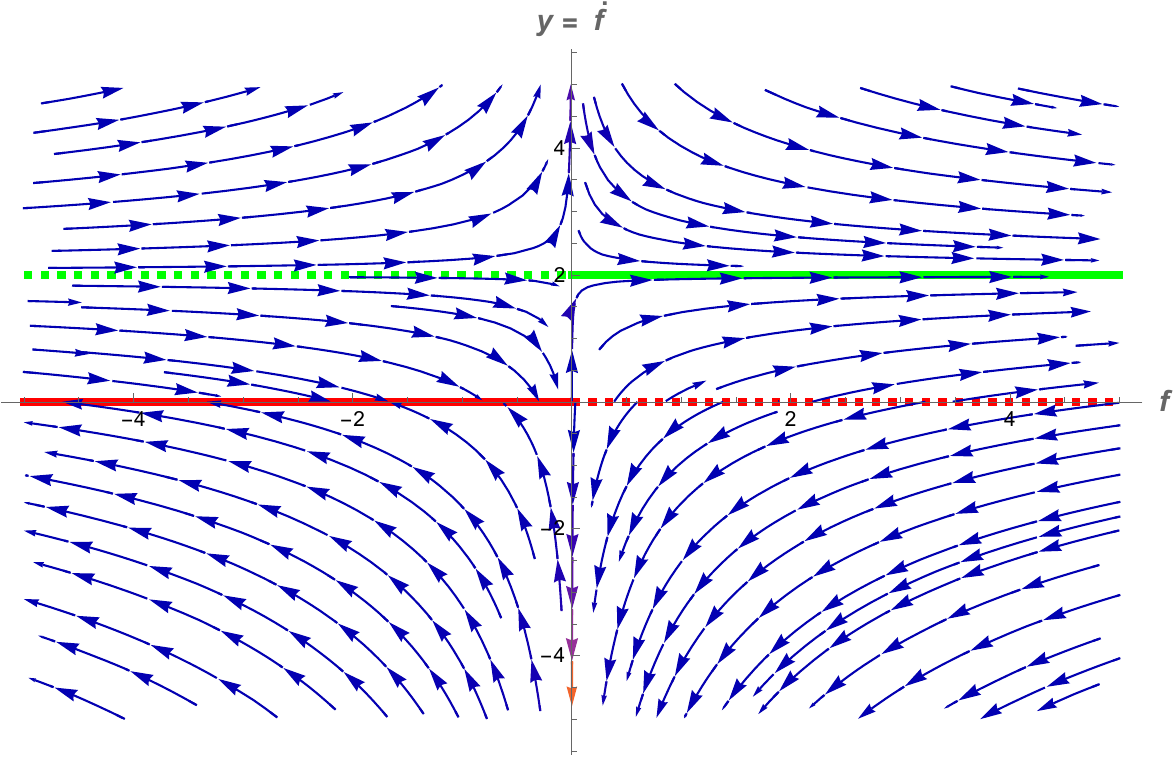}
\caption{Phase diagram for (\ref{bound1}) with $a=1$. The red line denotes the continuous line of fixed points with the solid half-line indicating the set of stable fixed points while the dashed half-line indicates the set of unstable fixed points. The solution is undefined along $f=0$. The nullcline is indicated in green with the solid portion indicating the attracting half-line while the dashed portion indicates the repelling half-line.} \label{pp1} 
\end{figure}

 The system (\ref{ds1}) possesses a continuous line of fixed points at $y_*=0$. The stability of these fixed points are usually determined by the determinant of the Jacobian of the right-hand-side of (\ref{ds1}) at $y_*=0$. Here, the determinant evaluates to zero which means that the fixed points are non-hyperbolic; one needs to be careful in extrapolating from the linear case to the fully nonlinear case based on this information. It is more instructive to plot the phase plane diagram of (\ref{ds1}). From Fig. \ref{pp1} it is clear that the continuous line of fixed points exhibit different behaviour depending on the sign of $f$: When $f<0$, then the half-line $y_*<0$ is stable (attracting), while, when $f>0$, the half-line $y_*>0$ is unstable (repelling). One obtains similar results when determining the phase plane via the level curves (\ref{firstint}).
 
 We note the existence of nullclines at $y=0$ (which is expected) as well as at $y=2a$. However, given the constraint (\ref{fdotcons}) we are only interested in the lower half plane. In fact, due to the fact that $f(t)\geq0$, we only consider the lower right quadrant. Here, we can see that the positive half-line of fixed points $f(t)\geq0$ is unstable. This means that trajectories move away from these points. The fixed points represent the constant initial static states that eventually collapse. Thus, at $t=0$, we have an initial static configuration. As $t$ increases, these states undergo unending collapse.

 \section{Charge and Cosmological constant effects}
  If we include the electromagnetic quantities $\ell(r)$ and $\eta$, where $\ell(r)$ represents the total charge within a sphere of radius $r$, and  $\eta$ is the proper charge density, and   the cosmological constant  $\Lambda$,  then the field equations become
\bs \label{ss3}
\bq
8 \pi \mu &=& \frac{3\dot{B}^2}{A^2B^2} - \frac{1}{B^2} \left(2\frac{B''}{B}-\frac{B'{}^2}{B^2} + \frac{4B'}{rB}\right)\nonumber \\
&&\mbox{}  -  \frac{\ell^2}{r^4B^4}-\Lambda, \label{ss3a} \\
8\pi p_r &=& \frac{1}{A^2}\left(-2\frac{\ddot{B}}{B}-\frac{\dot{B}^2}{B^2} + 2 \frac{\dot{A}\dot{B}}{AB}\right) \nonumber \\
&&\mbox{} + \frac{1}{B^2}\left(\frac{B'{}^2}{B^2}+2\frac{A'B'}{AB} + \frac{2A'}{rA}+\frac{2B'}{rB}\right)\nonumber \\
&&\mbox{} +  \frac{\ell^2}{r^4B^4}+\Lambda, \label{ss3b} \\
8\pi p_t &=& \frac{1}{A^2}\left(-2\frac{\ddot{B}}{B}-\frac{\dot{B}^2}{B^2} + 2 \frac{\dot{A}\dot{B}}{AB}\right)  \nonumber \\
&&\mbox{} + \frac{1}{B^2} \left( \frac{B''}{B} - \frac{B'{}^2}{B^2} + \frac{B'}{rB} + \frac{A''}{A} + \frac{A'}{rA}\right)   \nonumber \\
&&\mbox{} - \frac{\ell^2}{r^4B^4}+\Lambda, \label{ss3c} \\
8 \pi q &=& -\frac{2}{AB^2} \left(-\frac{\dot{B}'}{B}+\frac{B'\dot{B}}{B^2} + \frac{A'\dot{B}}{AB}\right), \label{ss3d} \\
4 \pi \eta &=& \frac{\ell'}{r^2B^3},\label{ss3e}
\eq \es
 which reduce to 
\bs \label{sss3}
\bq
8 \pi \mu &=&  \frac{1}{f^2}\left(8\pi \mu_0+\frac{3}{A_0^2}\dot{f}^2\right) -  \frac{\ell^2}{r^4B_0^4f^4}-\Lambda, \label{sss3a} \\
8\pi p_r &=& \frac{1}{f^2}\left(8\pi (p_r)_0 - \frac{1}{A_0^2}\left(2 f \ddot{f} +\dot{f}^2\right)\right) +  \frac{\ell^2}{r^4B_0^4f^4}+\Lambda, \label{sss3b} \\
8\pi p_t &=& \frac{1}{f^2}\left(8\pi (p_t)_0 - \frac{1}{A_0^2}\left(2 f \ddot{f} +\dot{f}^2\right)\right) - \frac{\ell^2}{r^4B_0^4f^4}+\Lambda, \label{sss3c} \\
8 \pi q &=&  -\frac{2A_0'}{A_0^2B_0^2}\frac{\dot{f}}{f^3}, \label{sss3d} \\
4 \pi \eta &=& \frac{\ell'}{r^2B_0^3f^3},\label{sss3e}
\eq \es
for the ansatz (\ref{pot}).  

The boundary condition (\ref{match}) now becomes (noting that $(p_r)_0=0$ on the stellar boundary)
\be 2 f \ddot{f} + \dot{f}^2 - 2 a \dot{f} - c \frac{1}{f^2} - d f^2 = 0, \label{bound2} \ee
where 
\be a = \frac{A_0'}{B_0} > 0, \qquad c=\frac{A_0^2\ell^2}{r^4 B_0^4}>0,  \qquad d=A_0^2 \Lambda,\label{constants} \ee
and the exterior is now the  charged Vaidya solution \cite{jaryl11}.

We note that the presence of charge reduces the density compared to the neutral case.  It also reduces the tangential pressure but increases the radial pressure. A positive (negative) cosmological constant  reduces (increases) the density but increases (decreases) both the radial and tangential pressure. The positivity of the radial pressure, taking (\ref{bound2}) into account, requires (\ref{fdotcons}) again. Thus,  an initially static stellar configuration must collapse; the presence of charge and/or a cosmological constant cannot effect expansion.

\subsection{Charged fluid with zero cosmological constant}
\begin{figure}[t]
\includegraphics[width=5in]{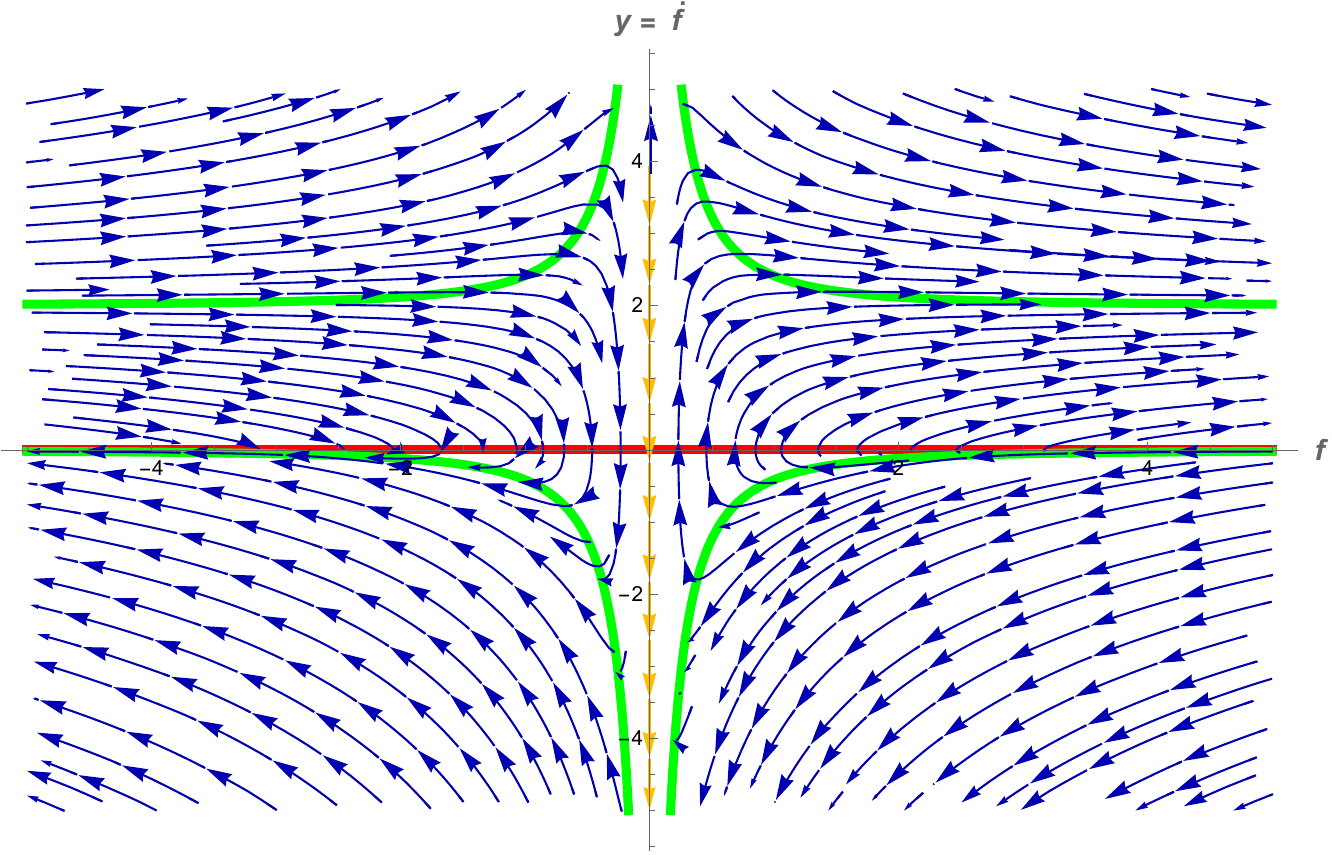}
\caption{Phase diagram for (\ref{bound2}) with $a=c=1$. The red line denotes the $f$-nullcline while the green lines denote the $y=\dot{f}$-nullclines. The solution is undefined along $f=0$.} \label{chargedfig} 
\end{figure}

When $\ell\neq0,\Lambda=0$, (\ref{bound2}) becomes
\be 2 f \ddot{f} + \dot{f}^2 - 2 a \dot{f} - c \frac{1}{f^2}  = 0, \label{bound3} \ee
which can be expressed as 
 \bs \label{ds33}
\bq
\dot{f}&=&y, \label{ds33a} \\
2 f \dot{y} &=& - y^2+2 a y + c \frac{1}{f^2}, \label{ds33b}
 \eq \es
 and we can see that the system 
has no fixed points. The phase portrait of (\ref{bound3}) is illustrated in Fig. \ref{chargedfig}. Again, we are interested in trajectories in the lower right quadrant. It would seem that the majority of trajectories approach $(0,-\infty)$. However, all trajectories will eventually intersect the nullcline and terminate on the $\dot{f}=0$ line. The end state is thus a superdense cold star. The presence of charge effects a braking effect on the unending  collapse of a neutral star seen in Fig. \ref{pp1}. 

\subsection{Neutral fluid with nonzero cosmological constant}

\begin{figure}[t]
\includegraphics[width=5in]{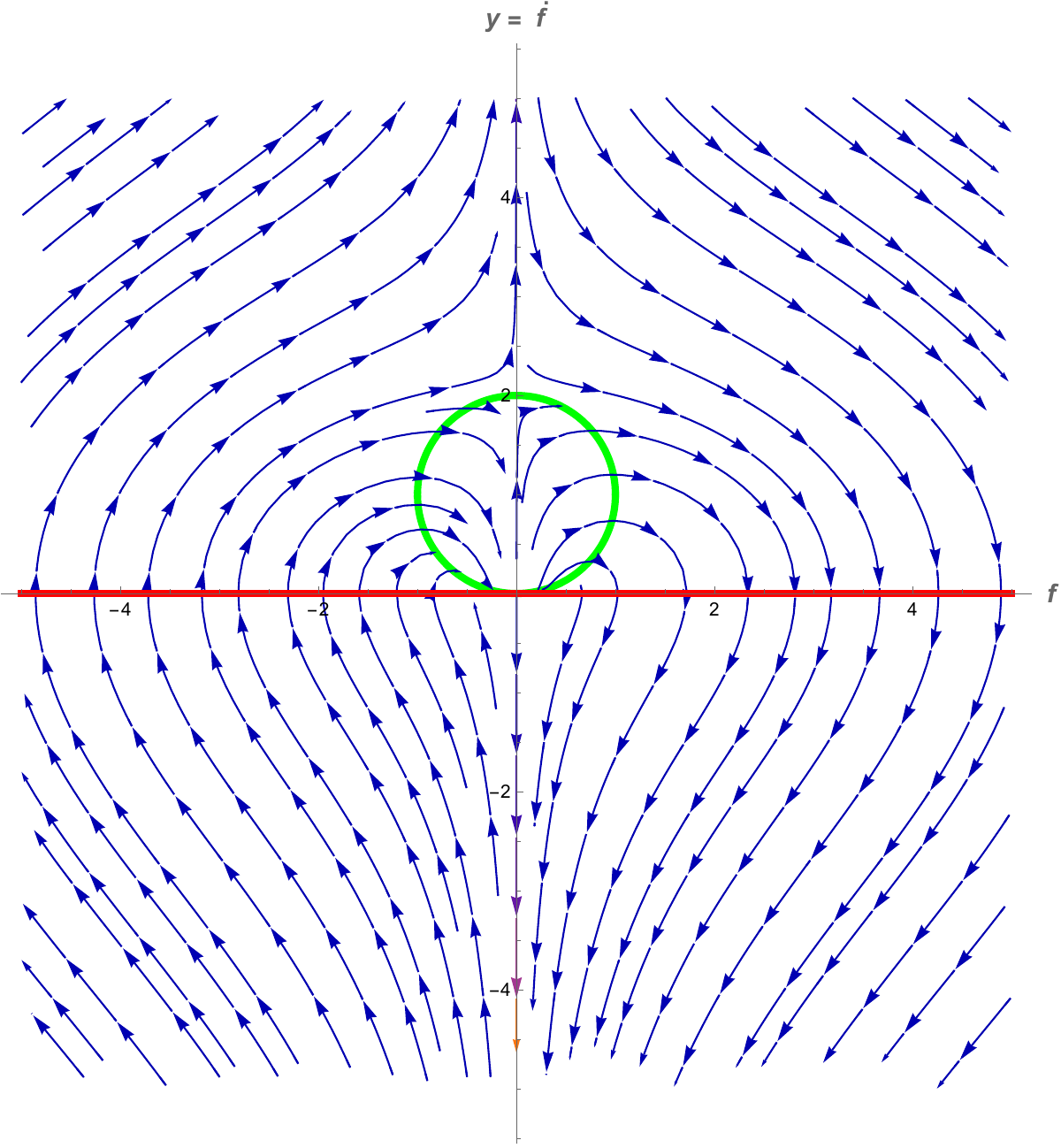}
\caption{Phase diagram for (\ref{bound2}) with $a=-d=1$. The red line denotes the $f$-nullcline while the green line denote the $y=\dot{f}$-nullclines. The solution is undefined along $f=0$.} \label{negcosmo} 
\end{figure}

\begin{figure}[t]
\includegraphics[width=5in]{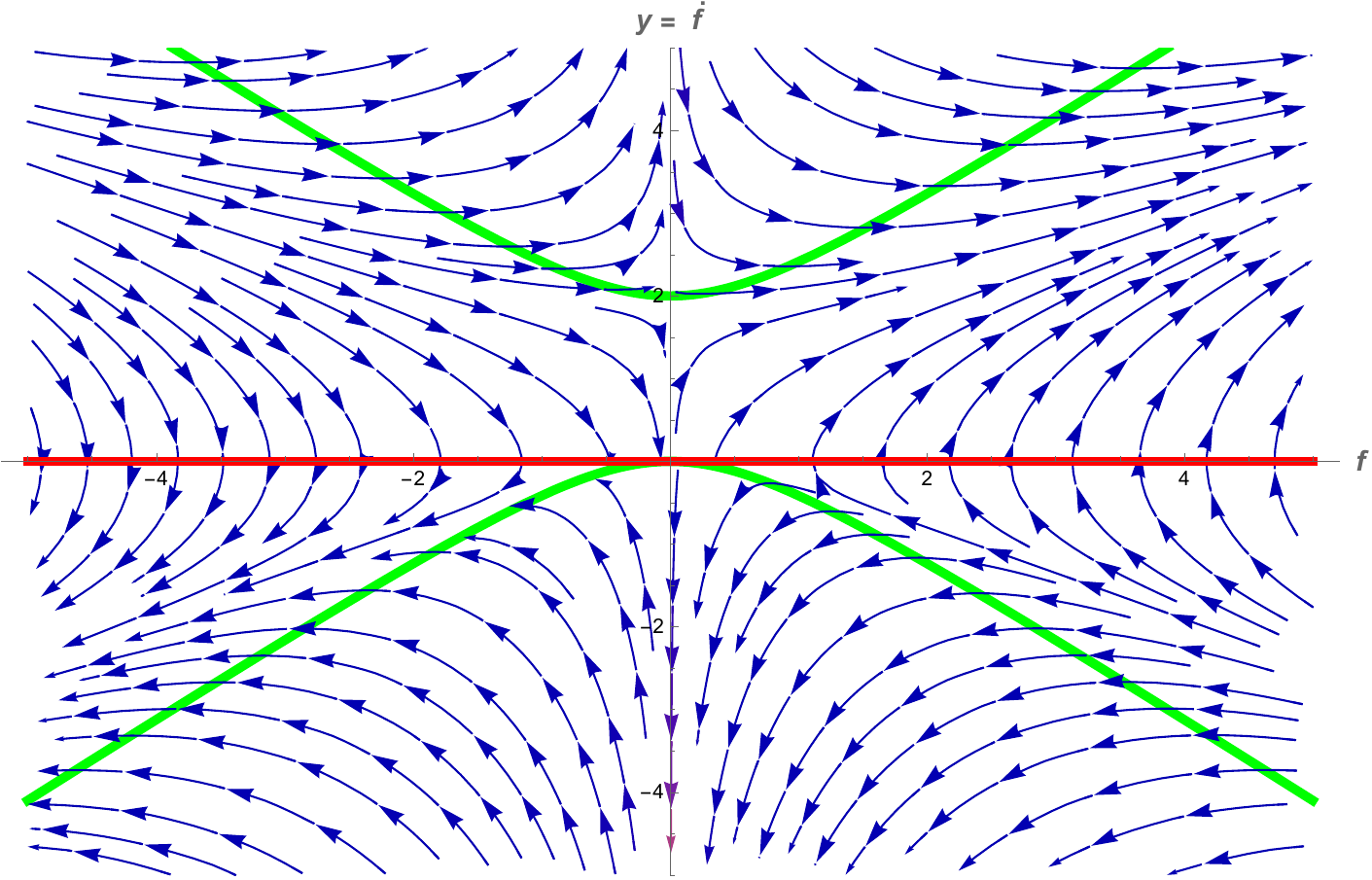}
\caption{Phase diagram for (\ref{bound2}) with $a=d=1$. The red line denotes the $f$-nullcline while the green lines denote the $y=\dot{f}$-nullclines. The solution is undefined along $f=0$.} \label{poscosmo} 
\end{figure}

 When $\ell=0,\Lambda\neq0$, (\ref{bound2}) becomes
\be 2 f \ddot{f} + \dot{f}^2 - 2 a \dot{f}  - d f^2 = 0, \label{bound4} \ee
which  can be written as the system
 \bs \label{ds22}
\bq
\dot{f}&=&y, \label{ds22a} \\
2 f \dot{y} &=& - y^2+2 a y +d f^2.\label{ds22b}
 \eq \es
 One can show that, while $f=0$ is a singularity, 
 \be (f_*,y_*)=(0,0), \label{singfp}\ee
 plays the role of a fixed point. Given that we are interested in the behaviour of the trajectories in the lower right quadrant $(f<0,\dot{f}>0)$, this does not have a significant impact on our analysis. Phase portraits for (\ref{ds22}) are given in Fig. \ref{negcosmo} and Fig. \ref{poscosmo} for negative and positive cosmological constant, respectively. In both diagrams, the nullclines are tangent at (\ref{singfp}) and the fixed point  is half-stable. In the case of a negative cosmological constant (Fig. \ref{negcosmo}), all trajectories in the lower right quadrant approach $(0,-\infty)$, signifying  unending collapse. No superdense cold stars can form in this scenario. On the other hand, when $\Lambda>0$ (Fig. \ref{poscosmo}), while some trajectories cross the nullcline and approach $(0,-\infty)$, a significant portion terminate on the $\dot{f}=0$ line indicating the possibility of superdense cold stars as the fate of stellar collapse. This separation of trajectories occurs along the stable manifold of (\ref{singfp}).

 \subsection{Charged fluid with nonzero cosmological constant}

\begin{figure}[t]
\includegraphics[width=5in]{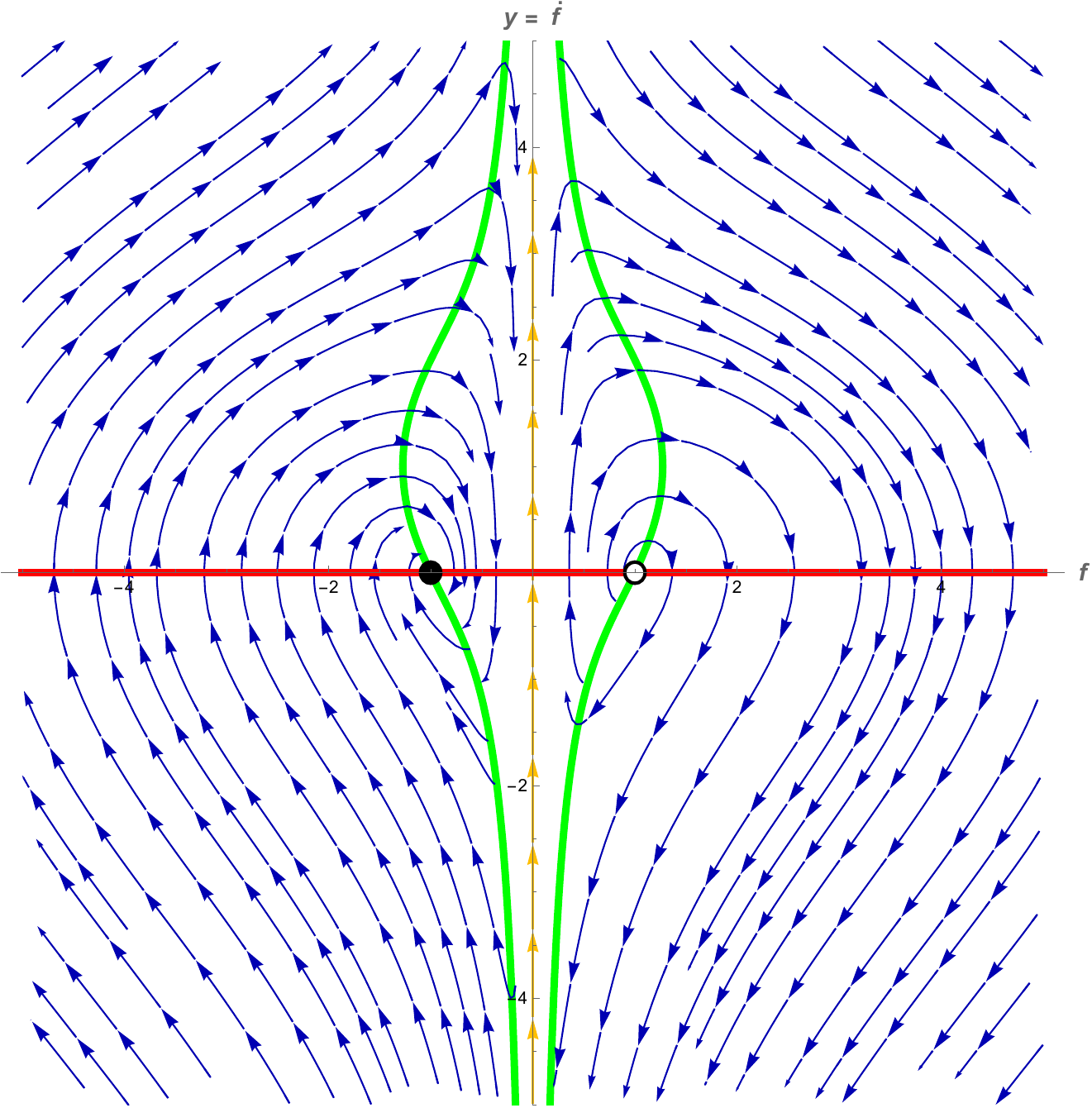}
\caption{Phase diagram for (\ref{bound2}) with $a=c=-d=1$. The red line denotes the $f$-nullcline while the green lines denote the $y=\dot{f}$-nullclines. The solution is undefined along $f=0$.} \label{gennegcosmo} 
\end{figure}

 When $\ell\neq0,\Lambda\neq0$, we rewrite (\ref{bound2}) as the system 
  \bs \label{ds2}
\bq
\dot{f}&=&y, \label{ds2a} \\
2 f \dot{y} &=& - y^2+2 a y +c\frac{1}{f^2}+ d f^2.\label{ds2b}
 \eq \es
This system has two fixed points given by 
\be (f_*, y_*) = \left(\pm \left(-\frac{c}{d}\right)^\frac14,0\right), \label{genfp} \ee
which exist provided $\Lambda<0$. Due to the positivity of the discriminant of the associated Jacobian evaluated at (\ref{genfp}) we need to look at the trace which is given by
\be \tau = \pm a \left(-\frac{d}{c}\right)^\frac14. \ee
We conclude that the positive fixed point is unstable while the negative fixed point is stable. The phase portrait in this case is given in Fig. \ref{gennegcosmo}. We observe that the trajectories in the lower right quadrant cross the nullcline and terminate on the 
 $\dot{f}=0$ line.  This braking effect on the collapse is clearly due to the presence of the charge as a similar phenomenon was absent in Fig. \ref{negcosmo}, where $\Lambda<0$.

\begin{figure}[t]
\includegraphics[width=5in]{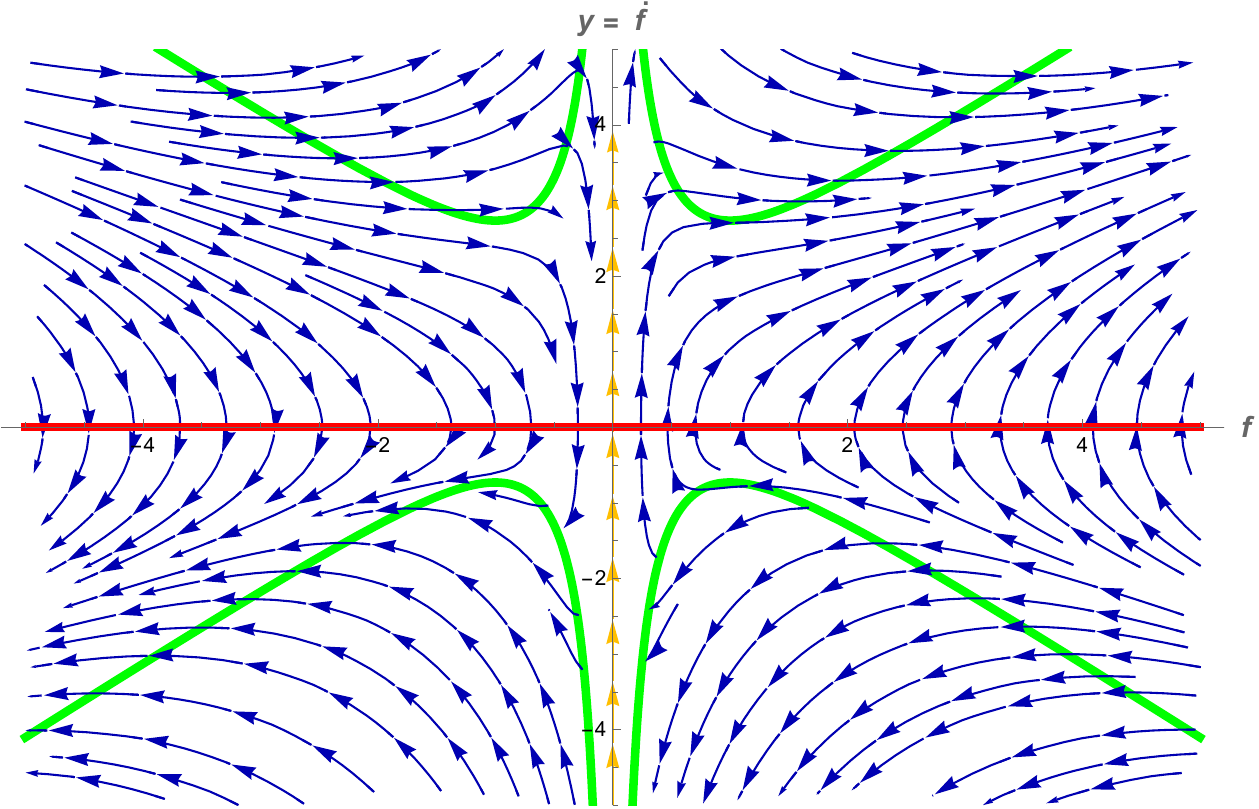}
\caption{Phase diagram for (\ref{bound2}) with $a=c=d=1$. The red line denotes the $f$-nullcline while the green lines denote the $y=\dot{f}$-nullclines. The solution is undefined along $f=0$.} \label{genposcosmo} 
\end{figure}

The phase portrait for positive $\Lambda$  is given in Fig. \ref{genposcosmo}. As in the neutral case (Fig. \ref{poscosmo}), one can observe trajectories which terminate on the $\dot{f}=0$ line. However, the trajectories that approach $(0,-\infty)$ in that case now cross the nullcline again and also terminate on the $\dot{f}=0$ line. 

\begin{figure}[t]
\includegraphics[width=5in]{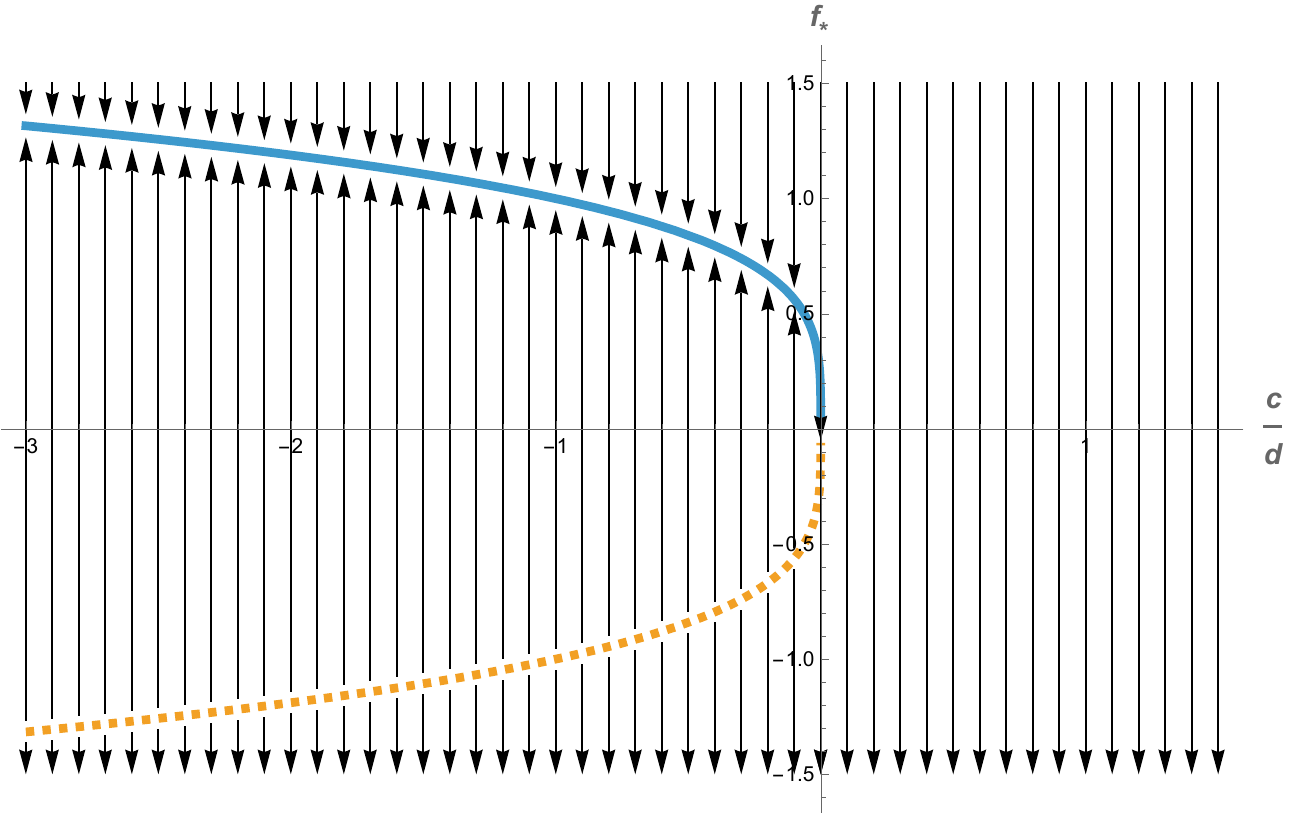}
\caption{Bifurcation diagram for (\ref{ds2}) indicating the presence of a fold bifurcation. The solid line denotes the stable positive fixed point while the dashed line denotes the negative unstable fixed point.} \label{bdiag} 
\end{figure}

 We observe that there is a fold bifurcation with $c/d$ (a combination of charge and the cosmological constant) playing the role of the bifurcation parameter (See Fig. \ref{bdiag}.). These bifurcations (also known as saddle-node, limit point, or turning-point bifurcations) are fundamental in physics because they describe how a system undergoes a sudden, irreversible transition as an external parameter changes. They represent the tipping points where equilibrium states of opposite stability are spontaneously created or destroyed. While this phenomenon is apparent here in the shear-free case, we note that it was also observed in the shearing collapse of a radiating star \cite{gm2026}.
 
 \section{Discussion}
 
 We have undertaken a qualitative analysis of the shear-free evolution of an initial static configuration. Our investigation  leads to the conclusion that {\it initial static configurations will necessarily collapse from initial time $t_0$ until time $t_h$, the time of formation of the event horizon}. If the initial static configuration is isotropic, then the stellar collapse remains isotropic for its duration. However, in the event that the initial static configuration is anisotropic, the collapse will be anisotropic.  The physical interpretations that arise are different from the earlier study of \cite{mg2025}, where the matter configuration was not initially static. Consequently new dynamical features arise in this analysis.
 
 Our analysis also included the effect of charge and the cosmological constant. Of these two quantities, we found that the charge had the most dramatic impact; it caused a `braking' effect on the stellar collapse and led to the formation of superdense cold stars as opposed to the unending collapse in the neutral cases. The presence of a negative cosmological constant on its own did not qualitatively affect the collapse. However, a positive cosmological constant allowed for both unending collapse as well as the formation of superdense cold stars.

 We note that spacetimes of the form (\ref{int1}) with (\ref{pot}) have received much attention over the years. In particular, de Oliveira {\it et al} \cite{doks1,doks} considered the shear-free collapse of an initial static configuration with metric functions given by (\ref{pot}). In their case, the initial static configuration was isotropic so their analysis only applied to isotropic collapse. Here, we showed that one could take a similar approach for anisotropic pressure. Other recent investigations for initially static models include the treatments of Charan {\it et al} \cite{y1}, Mesquita and da Silva \cite{y2}, Bogadi {\it et al} \cite{y3}, and Govender and Govender \cite{y4}.
 
\begin{figure}[t]
\centering\includegraphics[width=4in]{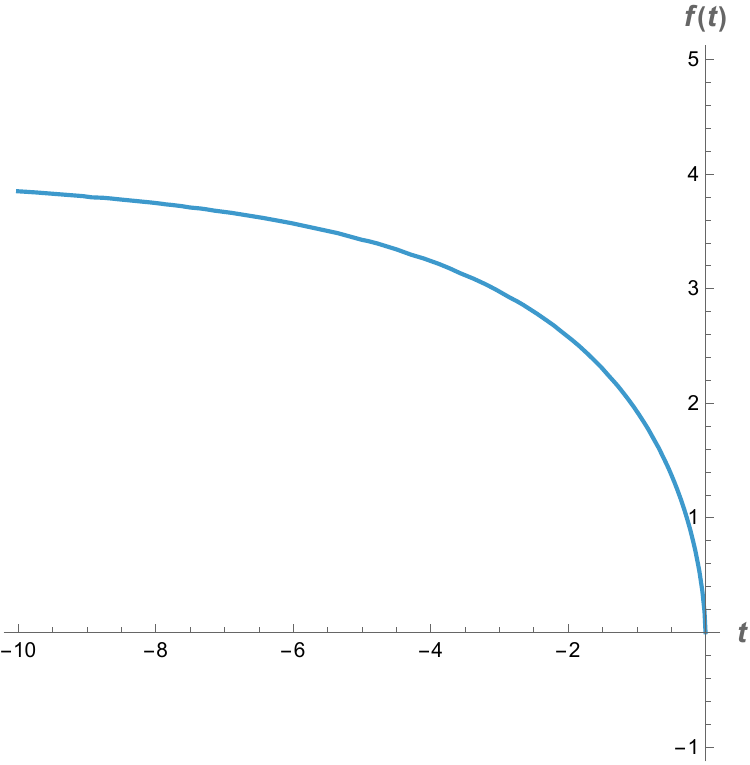}
\caption{Implicit plot of (\ref{impsol}) with $a=1$, $b=2$ and $c=0$. It is clear that the graph has a maximum value of 4 (or $b^2$) as $t\rightarrow-\infty$ and a minimum of zero at $t=0$.} \label{impsolfig} 
\end{figure}

 Our significant departure from earlier results is in the determination of the range for $t$.  Their analysis focussed on the implicit solution (\ref{impsol}) and noted that $f(t)$ in (\ref{impsol}) decreases monotonically from $b^2$ to zero as $t$ moves from $-\infty$ to zero. This led them to conclude that ``the collapse begins at $t=-\infty$'' \cite{doks1}. 
If we plot the solution (\ref{impsol}) we can indeed see that $t\in(-\infty,0)$ (Fig. \ref{impsolfig}). Thus, based on an analysis of (\ref{impsol}) as well as plots of the solution, the conclusion that $f(t)$ decreases monotonically from $b^2$ to zero as $t$ moves from $-\infty$ to zero seems reasonable. However, one has to be cautious in extrapolating that conclusion to the behaviour of all solutions of (\ref{bound1}). For example, consider the expression
 \be \sqrt{f} = t. \label{eg1} \ee
 If we analyse the behaviour of $f(t)$ based simply on the form (\ref{eg1}) we would conclude that $f>0$ and $t>0$. However, if we square both sides we obtain
 \be f = t^2 \label{eg2} \ee
 and then can conclude that, while we still have $f>0$, now $t\in(-\infty,\infty)$. The challenge with (\ref{impsol}) is that we cannot solve explicitly for $f(t)$ and so cannot determine immediately if our conclusion with respect to the behaviour of $f(t)$ and $t$ is correct. This is the main obstacle when finding exact solutions (implicit or otherwise) to differential equations: it is difficult to determine the qualitative behaviour of the solution even though the exact solution is known \cite{strog}. By undertaking a qualitative analysis of (\ref{bound1}) we have shown that $t$ does not need to be negative. Rather, it can evolve from any (positive) initial time.
  
As a final comment on the range of $t$, we note that (\ref{bound1}) admits a time translational invariance. As a result, one can change the initial value of $t$ by any additive quantity. Thus, adding an ``$\infty$'' to the range $(-\infty,0)$ moves it to $(0,\infty)$.

By undertaking a qualitative analysis of (\ref{bound1}) we have resolved the non-physicality of negative time introduced by other (quantitative) analyses.

\section*{Data Availability Statement}
 All data that support the findings of this study are included in the article (and any supplementary files).

\section*{Conflict of Interest}
We confirm that no conflicts of interest exist.

\section*{Acknowledgment}
KSG and SDM thank the University of KwaZulu--Natal for ongoing support.

\end{document}